\documentclass[twocolumn,showpacs,amsmath,amssymb,prl]{revtex4}

\usepackage{graphicx}
\usepackage{dcolumn}
\usepackage{bm}

\begin{document}

\preprint{APS/123-QED}

\title{Resonant Charge Relaxation as a Likely Source of the Enhanced Thermopower in FeSi}

\author{Peijie Sun$^{1}$}
\author{Beipei Wei$^{1}$}
\author{Dirk Menzel$^2$}
\author{Frank Steglich$^3$}
\affiliation{%
$^1$Beijing National Laboratory for Condensed Matter Physics, Institute of Physics, Chinese Academy of Sciences, Beijing 100190, China \\
$^2$Institut f\"ur Physik der Kondenierten Materie, Technische Universit\"at Braunschweig, D-38106 Braunschweig, Germany \\
$^3$Max Planck Institute for Chemical Physics of Solids, D-01187 Dresden, Germany
}%

\date{\today}

\begin{abstract}
The enhanced thermopower of the correlated semiconductor FeSi is found to be robust against the sign of the relevant charge carriers.
At $T$\,$\approx$\,70 K, the position of both the high-temperature shoulder of the thermopower peak and the nonmagnetic-enhanced paramagnetic crossover, the Nernst coefficient $\nu$ assumes a large maximum and the Hall mobility $\mu _H$ diminishes to below 1\,cm$^2$/Vs. These cause the dimension-less ratio $\nu$/$\mu _H$$-$a measure of the energy dispersion of the charge scattering time $\tau(\epsilon)$$-$to exceed that of classical metals and semiconductors by two orders of magnitude. Concomitantly, the resistivity exhibits a hump and the magnetoresistance changes its sign. Our observations hint at a resonant scattering of the charge carriers at the magnetic crossover, imposing strong constraints on the microscopic interpretation of the robust thermopower enhancement in FeSi.

\end{abstract}

\pacs{Valid PACS appear here}
\maketitle

FeSi is a prototypical correlated semiconductor, with its band gap captured, while overestimated, by density functional theory, essentially different from a Mott insulator. So far, it has been considered either a Kondo insulator \cite{fiskKI} due to the hybridization of a local state and conduction band, a nearly magnetic semiconductor \cite{SCR} based on spin fluctuation theory of itinerant electrons, or a correlated band semiconductor \cite{kunes} emphasizing local correlation effects in a band semiconductor. Electronic transport of FeSi is of particular interest, as illustrated by its large positive thermopower peak amounting to $S$\,$\geq$\,500\,$\mu$V/K at $T$$\approx$\,50\,K \cite{1stS,sales}, quick metallization above room temperature \cite{sales,Delaire}, and unusual magnetoresistance as well as Hall conductance induced by doping \cite{manyala}. The potential application of FeSi as a cryogenic thermoelectric (TE) material \cite{goodTE} further propels the ongoing debate on this compound.

In order to account for the large thermopower peak observed for FeSi, the phonon-drag effect \cite{1stS,sales}, an appropriate hole doping in conjuction with the narrow-gap and narrow-band features \cite{Jarl,Saso,Busch, sales2011}, and strong Hubbard correlations \cite{sluch} had ever been invoked. Using dynamical mean field theory (DMFT), Tomczak $et\,al.$ \cite{Tomczak} have recently identified a correlation-induced incoherence in FeSi, which is argued to be the driving force of the metallization as well as a variety of unusual physical properties including the thermopower near the metallization crossover. This is to be compared to recent inelastic neutron scattering results \cite{Delaire}, which suggested an enhanced thermal disorder to account for the metallization. It is fare to say that a consistent interpretation of the large thermopower peak in FeSi, which is located at a temperature well below the metallization crossover, is still far from being reached.

A careful examination of experimental data in the literature reveals that in contrast to the Hall coefficient ($R_H$), the thermopower of FeSi is quite robust against variation of composition and crystallinity. As shown in Fig.\,1, while the $S(T)$ peak persists to be large and positive for all investigated FeSi samples, $R_H(T)$ shows a strong sample dependence, pointing to competing bands with either hole \cite{1stS,paschen, sluch} or electron-like \cite{1stS,sales} character dominating in the temperature range of interest, 10$-$100\,K (cf. inset of Fig. 1). These observations strongly hint at an unconventional origin of the enhanced thermopower in FeSi, contrasting the aforementioned phonon-drag and hole-doping scenarios, which predict the same sign of $S(T)$  and $R_H(T)$ in a substantial temperature range.

\begin{figure}[t]
\includegraphics[width=0.95\linewidth]{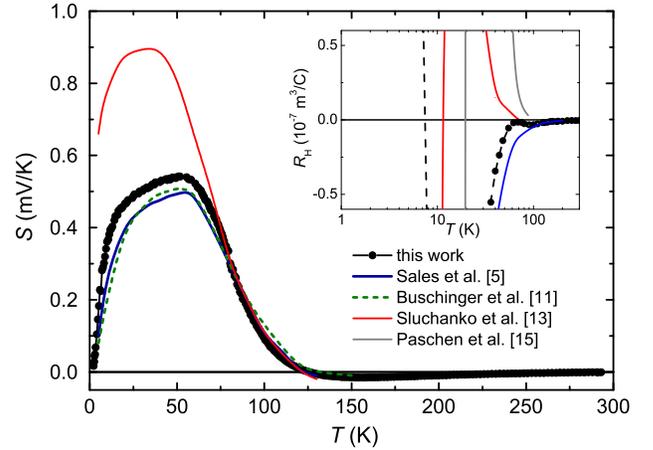}
\caption{Thermopower $S(T)$ (main panel) and Hall coefficient $R_H(T)$ (inset) of FeSi in comparison with literature data \cite{sales,sluch,paschen,Busch}. Note that, while $R_H(T)$ is extremely sample dependent showing either sign in the temperature range of interest (10$-$100 K), $S(T)$ is rather robust against the variation of the sign of the relevant charge carriers: it always exhibits a positive peak between 30 and 50 K, albeit of varying height.
\label{nus.eps}}
\end{figure}

Here, we report results of extensive magneto-transport measurements including the Nernst effect, Hall effect, magnetoresistance (MR) and thermopower, performed on a high-quality FeSi single crystal grown by the tri-arc Czochralski technique.  Upon decreasing temperature, a number of unusual transport phenomena, e.g., a sign change of MR($T$), a diminishing Hall mobility $\mu_H(T)$ (= $R_H$/$\rho$) and an unexpectedly large Nernst coefficient $\nu(T)$ concur at $T$\,$\approx$\,70\,K, i.e., at the high-temperature shoulder of the thermopower peak. Surprisingly, the magnitude of the latter two quantities is {\it anti-correlated} to each other, opposite to the common expectation of both being proportional to each other \cite{behnia1}. As we will discuss below, this signals a resonant charge relaxation process, presumably inherent to the crossover from both nonmagnetism to enhanced paramagnetism and coherent to incoherent electronic excitations \cite{Tomczak}.

In Fig.\,2 we show the complete set of the measured transport coefficients. Four characteristic temperatures from $T_1$ to $T_4$ are marked on top of the figure to facilitate our discussion. The thermopower $S$ and Nernst coefficient $\nu$ (cf.\,Fig.\,2(a)) were determined following the respective definition $S$\,=\,$E_x/|\Delta T_x|$ and $\nu$\,=\,$E_y/B_z |\Delta T_x|$, with $\Delta T_x$ being the applied temperature gradient, $B_z$\,=\,2\,T the magnetic field and $E_x$($E_y$) the induced electrical potential along the $x$\,($y$) direction. As has already been pointed out, opposite to the positive $S(T)$ values observed in the temperature range 2$-$120 K, our sample shows negative values of $R_H(T)$ in the whole temperature range investigated, except for $T$\,$<$\,$T_1$\,$=$\,8\,K [cf. Fig.\,2(b)]. Such a disparity in the signs of $S(T)$ and $R_H(T)$ might be ascribed to multiband effects. However, this is unlikely to address to the current FeSi sample: Significant non-linearity in the Hall resistivity, $\rho _H(B)$, is restricted only to $T$\,$<$\,10\,K (cf. inset of Fig.\,2), which indicates that multiband competition in the temperature range of interest is of minor importance. Rather, given the correlated nature of FeSi, electron-electron correlations appear to be relevant for the robust thermopower enhancement, as has been discussed for NaCo$_2$O$_4$ \cite{koshibae}.

\begin{figure}[t]
\includegraphics[width=1\linewidth]{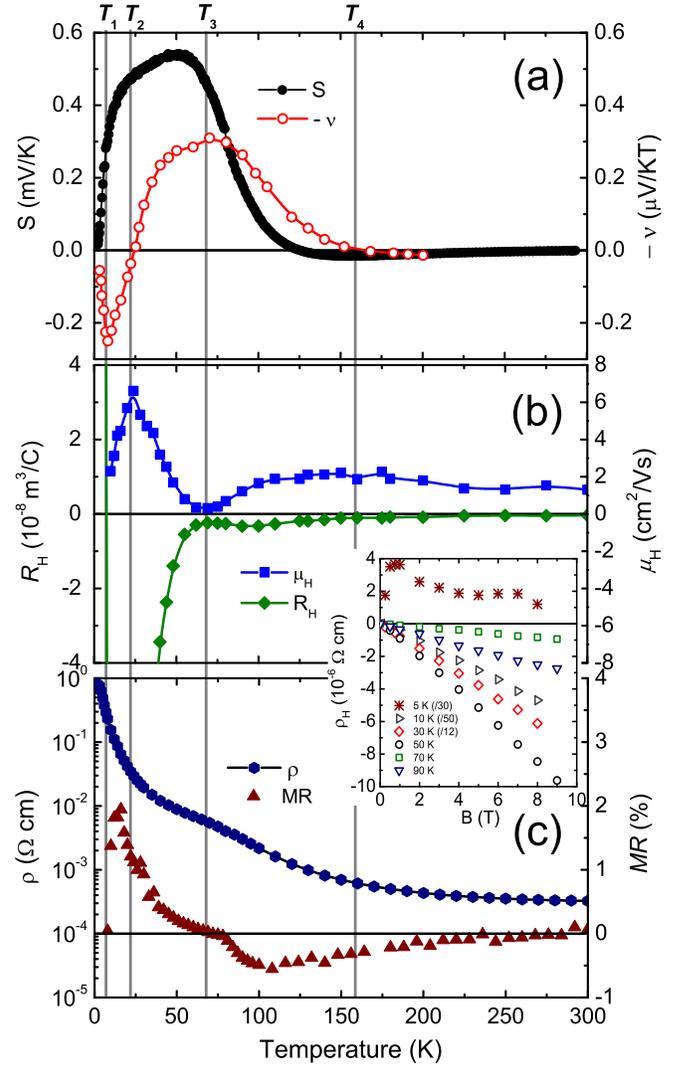}
\caption{Various transport coefficients for FeSi as a function of temperature. (a) Thermopower $S(T)$ and Nernst coefficient $\nu(T)$; (b) Hall coefficient $R_H(T)$ and Hall mobility $\mu_H(T)$; (c) resistivity $\rho(T)$ and magnetoresistance MR$(T)$, the latter being measured in 8\,T. $\mu_H(T)$ is shown only for $T$\,$>$\,10\,K, where the one-band approximation is applicable.
Vertical lines represent four characteristic temperatures as discussed in text.
Inset: Hall resistivity $\rho_H$ vs magnetic field $B$ at selected temperatures.
\label{nus.eps}}
\end{figure}

Corresponding to the notable thermopower peak at 50\,K, the value of $-\nu(T)$ is enhanced below $T_4$\,$\approx$\,160\,K, where it undergoes a sign change, and gradually develops a maximum at $T_3$\,=\,70\,K. $T_3$ is at the vicinity of the nonmagnetism-paramagnetism crossover \cite{fiskKI} and is of particular importance for FeSi: Here the maximum of $-\nu(T)$ concurs with the minimum of $\mu_H(T)$, the sign change of MR($T$), and the hump in $\rho(T)$ [cf. Fig.\,2(c)]. With further decreasing temperature, $-\nu(T)$ changes its sign again at $T_2$\,=\,23\,K where $\mu_H$ develops a maximum. Lowering the temperature further more leads to a negative peak of $-\nu(T)$ at around $T_1$\,=\,8\,K, where MR changes sign again. Interestingly enough, the two extrema of $-\nu(T)$ at $T_1$ and $T_3$ concur with the consecutive sign changes of MR($T$) and as least one minimum in $\mu_H(T)$ \cite{note2}; the two sign changes of $\nu(T)$ (at $T_2$ and $T_4$) meet the maxima in $\mu_H(T)$. These observations provide strong evidences for the Nernst effect being a sensitive probe, reflecting how the charge carriers are relaxed in different temperature regions.

In the past decade, intensive investigations of the Nernst effect have revealed important new insights into the physics of high-$T_c$ superconductors  \cite{wang} and magnetic semiconductors \cite{Pu}. In a nonmagnetic, nonsuperconducting compound, $\nu(T)$ is known to be very small, determined by the energy asymmetry of the charge relaxation spectra, as was originally described by Sondheimer \cite{Sonder, behnia1}. Additional sources for Nernst response, beyond Sondheimer's theorem, may arise from a distorted electronic structure, i.\,e., multiband effects, anisotropies or low dimensionality \cite{Clayhold}. However, in the case of FeSi, its simple cubic B20 structure, the absence of superconductivity and magnetic ordering, as well as the fact that the electronic transport properties are well described within the one-band approximation in a substantial temperature range, exclude all the afore-mentioned possibilities, leaving an anomalous charge relaxation spectrum as the most likely cause producing the enhanced $\nu(T)$ that is {\it anti-correlated} to $\mu_H(T)$. The excellent correspondence of the positions of the extrema and sign changes in $\nu(T)$ with those in MR($T$) and/or $\mu_H($T$)$ curves lend strong support to this proposition.

Starting from Sondheimer's description \cite{Sonder}, $\nu$ of a single, degenerate electronic band is expressed as the energy derivative of the Hall angle tan$\theta _H$ (= $\sigma _{xy}$/$\sigma _{xx}$) at the Fermi energy $\epsilon _F$. In the low field limit ($\mu_H$$B$\,$\ll$\,1), tan$\theta _H$ can be simply represented by $\mu _H$ and the relaxation time $\tau$: tan$\theta _H = \mu _H B = eB\tau/m^*$, with $m^*$ being the effective mass of the charge carriers. Assuming a typical power-law dependence of $\tau$\,$\approx$\,$\tau _0\,\epsilon^r$ \cite{Seeger},  $\nu$ is straightforwardly related to $\mu _H$ \cite{behnia1} (in ref.\,16, $r$ was assumed to be 1),
\begin{equation}
\nu\,=\,-\frac{\pi^2}{3}\frac{k_BT}{\epsilon _F}\frac{k_B}{e} \mu _H r.
\end{equation}
For a nondegenerate system, Eq.\,1 otherwise reduces to \cite{Seeger},
\begin{equation}
\nu\,=\,-\frac{k_B}{e} \mu _H r.
\end{equation}
Note that Eqs.\,1 and 2 are equivalent at the boundary of the degenerate and nondegenerate statistics, where $\epsilon _F$\,=\,$(\pi^2/3)k_BT$.
Typical charge scattering processes include the ones by acoustic phonones with $r$\,=\,$-1/2$, by polar optical phonons with 0\,$<$\,$r$\,$<$\,$1/2$,  and by ionized impurities with $r$\,=\,$3/2$ \cite{Seeger}.  All of them are of the order of unity, implying a weak energy dependence of $\tau(\epsilon)$.

Equations 1 and 2 state that in a single-band solid, $\nu(T)$ is expected to be {\it proportional} to $\mu_H(T)$ given that $r$\,$=$\,const. Such a correlation has indeed been verified for various compounds, including heavy-fermion (HF) systems \cite{behnia1} in the zero-temperature limit, where the enhancement of $\nu(T)$ is due to the greatly reduced $\epsilon _F$ (cf.\,Eq.\,1), or equivalently, the largely enhanced $m^*$. As already mentioned, in FeSi, we observe that the absolute values of $\nu(T)$ (Fig.\,2(a)) and $\mu_H(T)$ (Fig.\,2(b)) to be {\it anti-correlated} to each other: For example, the maxima of $|\nu(T)|$ concur with the minima of $\mu_H(T)$, whereas at the temperatures where $\nu(T)$ crosses zero, $\mu_H(T)$ assumes a maximum. To our knowledge, such phenomena have so far never been reported for any systems.

\begin{figure}[tp]
\includegraphics[width=0.99\linewidth]{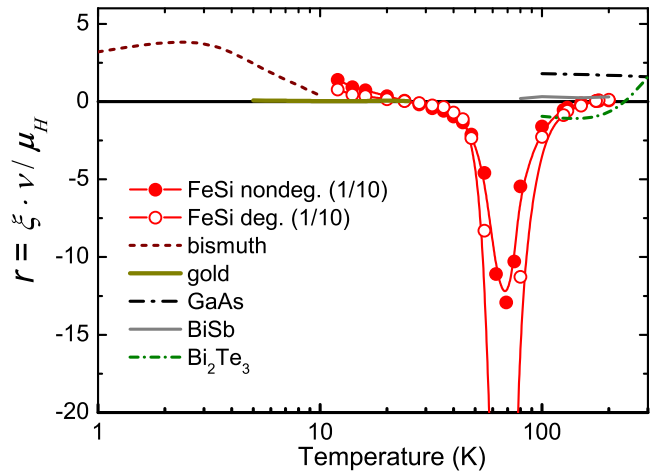}
\caption{Dimensionless ratio of $r$\,=\,$\xi$\,$\cdot$\,$\nu /\mu _H$ as a function of temperature for FeSi, compared with those of bismuth \cite{behnia1}, gold \cite{behniaAu}, GaAs (sample 9n) \cite{scatter.r}), BiSb \cite{BiSb}, and Bi$_2$Te$_3$ (sample R8) \cite{BiTe}). The prefactor $\xi$ is employed to keep the ratio dimensionless: It equals $-(\frac{\pi^2}{3}\frac{k_BT}{\epsilon _F}\frac{k_B}{e})^{-1}$ and $-(\frac{k_B}{e})^{-1}$ in the degenerate (bismuth and gold) and nondegenerate (GaAs, BiSb and Bi$_2$Te$_3$) approximation \cite{note}, respectively (cf.\,Eqs.\,1 and 2). Note that the ratio for FeSi was obtained within both approximations, and is plotted here divided by a factor of 10.
\label{rfactor.eps}}
\end{figure}

The validation of the single-band approximation enables one to estimate the scattering exponent $r$ simply by computing the dimensionless ratio of $\nu$ to $\mu_H$, cf. Eqs.\,1-2. This procedure ignores the fact that $\tau(\epsilon)$ might be complex due to a combination of different scattering mechanisms in, e.g., the vicinity of the sign changes and maxima of $\nu(T)$. For simplicity, we assume one prevailing dispersion relation of the charge relaxation at all temperatures. As shown in Fig.\,3, referring to the literature values of $\nu$ and $\mu_H$ for various classical semiconducting or metallic materials, the dimensionless ratio $r$\,$=$\,$\xi$$\cdot$$\nu$/$\mu_H$ are all estimated to be in the vicinity of unity, reflecting the ``classical" scattering processes off, e.\,g., acoustic phonons ($\xi$ is a prefactor eliminating the dimension of the ratio, see the caption of Fig.\,3).
By contrast, the ratio $r$ for FeSi, estimated within both degenerate and non-degenerate approximation \cite{note}, exhibits enormous values at around $T_3$, exceeding the ones in ordinary solids by two orders of magnitude. Note that the ratio for bismuth is not huge, as opposed to the giant value of $\nu$\,$\approx$\,7\,mV/KT at $T$\,$\approx$\,4\,K \cite{behnia1}. Indeed, this enhanced low-$T$ Nernst coefficient of bismuth had been attributed to a huge $\mu _H$ and a tiny $\epsilon _F$, in accord with the prediction made for simple degenerate systems by Eq.\,1. The somewhat larger ratios, relative to unity, for bismuth can be ascribed to the ambipolar effect, as is anticipated for a compensated semimetal.

The huge dimensionless ratio of $\nu$ to $\mu_H$ evidences a failure of the anticipated power-law dependence of $\tau(\epsilon)$, hinting at a resonant charge relaxation in the vicinity of the thermopower peak in FeSi. Notably, this feature is concomitant to the crossover from nonmagnetism to enhanced paramagnetism of FeSi \cite{fiskKI}. This suggests an involvement of the ``unlocked'' magnetic moment, which is likely caused by strongly enhanced magnetic fluctuations in FeSi \cite{Tomczak}, mimicking the case of a Kondo-resonance scattering. Further support for this proposition is lent from the magnetoresistance MR($T$), which exhibits a sign change exactly at the temperature of the magnetic crossover ($T_3$). The negative MR above $T_3$ can reasonably be ascribed to the spin-related scattering mechanism of charge carriers. By contrast, the second sign change of MR($T$) concuring with another extremum of $|\nu(T)|$ at $T_1$ has already been reported \cite{MRFeSi}, and the negative MR below it is presumably due to quantum interference effects \cite{paschen}.

The striking experimental evidence for a highly dispersive $\tau(\epsilon)$ at around $T_3$ will help us to argue below that it is intimately related to the robust thermopower enhancement in FeSi. This is obvious in a degenerate electron system, where the Mott expression for the thermopower states that $S$ measures the energy-dependent electrical conductivity, which is determined by the dispersive electronic density of states (DOS) $N(\epsilon)$ and the dispersive scattering time $\tau(\epsilon)$ at $\epsilon_F$ \cite{goodTE,sun12}. The first term of dispersive $N(\epsilon)$ dominates the thermopower in the majority of conducting degenerate materials, explaining the frequently observed equality of the signs of $S(T)$ and $R_H(T)$. In contrast, two of the authors have shown that in a prototypical Kondo-lattice system, the dispersive $\tau(\epsilon)$ of the conduction electrons, derived from the Kondo scattering, dominates the enhanced thermopower over a surprisingly wide temperature range \cite{sun12}.
Similarly, for a nondegenerate semiconductor, the dispersive $\tau(\epsilon)$ turns out to be an important ingredient entering into the thermopower, in addition to the thermally activated charge carriers across the band gap $E_g$ \cite{sun13}.

Different to the thermopower originating from a dispersive $N(\epsilon)$, the sign of $S(T)$ due to a dispersive $\tau(\epsilon)$ is not bound to that of the relevant charge carriers \cite{behnia1,sun12}: $\partial{\tau}(\epsilon)/\partial\epsilon$ can be of either sign for a certain type of charge carriers, opposite to the case of $\partial{N}(\epsilon)/\partial\epsilon$, whose sign is predetermined by the type of charge carriers.
Such considerations open a way to capture the positive thermopower peak being robust against the sign of $R_H(T)$. A substantial contribution to thermopower due to a resonant $\tau(\epsilon)$ in FeSi is expected at about $T_3$. While the Nernst coefficient is still significant at $T$\,$<$\,20\,K and assumes an extremum at $T_1$\,=\,8\,K, the corresponding dimensionless ratio $\nu$/$\mu_H$ in this lower-temperature range is obviously not large. Here, different charge relaxation processes (e.g., the one involved in quantum interference effects) with differing dispersion relations or the ambipolar effect (note that, $\rho_H(B)$ exhibits significant nonlinearity and $R_H$ changes sign at around $T_1$) may be entangled and invalidate our discussion based on a single band and a single relaxation process.

With a proper theoretical description lacking, the resonant charge-carrier relaxation inferred for FeSi near $T_3$ (\,$=$\,70\,K) is tentatively attributed to some significant many-body process that concurs with the nonmagnetic-paramagnetic crossover. The involvement of local magnetic moment in this process is reminiscent of the Kondo scattering of conduction electrons from localized magnetic moments, suggesting that $T_3$ may play the role of the Kondo temperature $T_{\rm K}$. Further support to this resonant relaxation scenario comes from the recent results of DMFT calculations, revealing a crossover from low-temperature coherent electronic excitations to high-temperature incoherent ones to take place at around $T_3$ \cite{Tomczak}. Indeed, such a crossover is reminiscent of the one between incoherent, local moment and coherent, HF regime in a Kondo lattice, where a thermopower maximum genetically occurs \cite{Zlatic05}. Moreover, it is interesting to note that the opposite signs of $S(T)$ and $R_H(T)$ as observed for FeSi, are also found in various doped Mott insulators \cite{koshibae, uchida11}. Despite all the indications of the involvement of electron-electron correlations described in this paper, a microscopic interpretation of the observed novel relaxation process appropriate for a $d$-electron system, implying the concurrence of electronic and magnetic crossovers, remains challenging.

We thank J. M. Tomczak, W. Xu, and G. Kotliar for stimulating discussions. P.S. thanks financial support from the MOST of China (Grant No: 2012CB921701) and the Chinese Academy of Sciences through the strategic priority research program (Grant No: XDB07020200). F.S. acknowledges partial support by the DFG through FG 960 ``Quantum Phase Transitions".


\begin{thebibliography}{99}
\bibitem{note2} Because $\mu_H$ is defined only for above 10\,K in order to avoid the multiband effects, we couldn't confirm a minimum of $\mu_H(T)$ at $T_1$.
\bibitem{note} Because the charge gap in FeSi opens at a rather low temperature of about $\sim$\,200\,K, either the degenerate or nondegenerate approximation seems not adequate for the statistics of the charge carriers.
\end{thebibliography}
\end{document}